\documentstyle[11pt,newpasp,epsf]{article}
\begin{document}
\title{Baryonic Tully-Fisher Relations}
\author{Stacy S. McGaugh}
\affil{Department of Astronomy, University of Maryland}
\begin{abstract}

I describe the disk mass--rotation velocity relation which underpins
the familiar luminosity--linewidth relation.  Continuity of this relation
favors nearly maximal stellar mass-to-light ratios.  This contradicts
the low mass-to-light ratios implied by the lack of surface brightness
dependence in the same relation.

\end{abstract}

\keywords{Dark Matter, Galaxies, Galaxy Formation, Galaxy Evolution,
Stellar Populations}

\section{Searching for the Physical Basis of the Tully-Fisher Relation}

The Tully-Fisher (TF) relation (Tully \& Fisher 1977) is well known.
Yet why it works is not clear. A dizzying variety of
distinct interpretations have been offered over the years
(e.g., Aaronson, Mould, \& Huchra 1979; Milgrom 1983;
Walker 1999).  There is no consensus even when the
context is limited to that of NFW halos
Widely divergent pictures have been offered,
sometimes in successive papers by the same
authors (e.g., Dalcanton, Spergel, \& Summers 1995, 1997; 
van den Bosch \& Dalcanton 2000;
Mo, Mao \& White 1998; Mo \& Mao 2000; Steinmetz \& Navarro 1999,
Navarro \& Steinmetz 2000).

It is commonly {\it assumed\/} that mass
scales with some power of rotation velocity, and that luminosity traces mass.
The first piece of this common wisdom is questionable given the startling
lack of dependence of the TF relation on surface brightness
(Sprayberry et al.\ 1995; Zwaan et al.\ 1995).
It matters not at all whether the luminous mass is concentrated or
diffuse.  This is commonly interpreted to mean that the mass in stars
is insignificant.  If stellar mass contributes noticeably to the
rotation velocity, $V^2 = G{\cal M}/R$ surely
demands {\it some\/} shift (McGaugh \& de Blok 1998; Courteau \& Rix 1999).

Whether luminosity traces mass is a more tractable issue.
I address this here in an empirical way using data which span the largest
available dynamic range.  This at least makes clear that the
fundamental relation which needs explaining is one between {\it rotation
velocity\/} and {\it disk mass\/} (McGaugh et al.\ 2000).

\section{The Disk Mass--Rotation Velocity Relation}

Implicit in our presumption that light traces mass is the relation
\begin{equation}
L = \Upsilon_*^{-1} f_* f_d f_b {\cal M}_{tot},
\end {equation}
where $f_b$ is the baryon fraction of the universe,
$f_d$ is the fraction of the baryons associated with a particular galaxy
which reside in the disk, $f_*$ is the fraction of disk baryons in the
form of stars, and $\Upsilon_*$ is the mass-to-light ratio of the stars. 
Each of the pieces which intervene between $L$ and
${\cal M}_{tot}$ must be a nearly universal constant shared by all disks, or
a finely tuned function of rotation velocity, in
order to maintain the observed TF relation.

We can improve on equation (1) by using the observed gas mass to correct
for $f_*$ (Fig.\ 1).  The TF relation works in bright galaxies because
they are star dominated: $f_* \ge 0.8$.  This breaks down as one examines
lower mass galaxies which are progressively more gas dominated
(McGaugh \& de Blok 1997).  Yet if we add in the gas mass, the TF relation
is restored (Fig.\ 1c).

\begin{figure}
\plotone{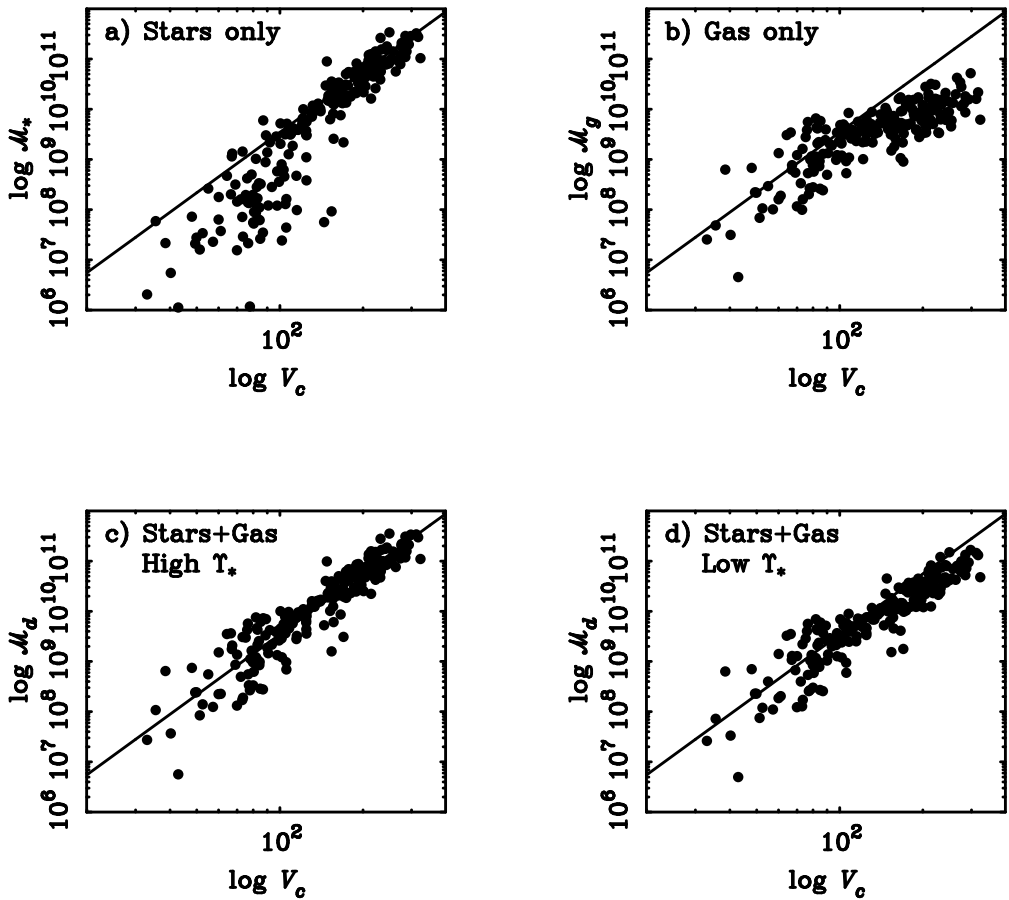}
\caption{TF relations constructed from the $H$-band data of
Bothun et al.\ (1985) and the $I$-band data of Pildis, Schombert, \& Eder
(1997).  The data of Bothun et al.\ (1985) cover the range $V_c > 100\;
{\rm km}\,{\rm s}^{-1}$ traditionally covered by TF studies.  The range
$V_c < 100\;{\rm km}\,{\rm s}^{-1}$ is now probed by data
from Eder \& Schombert (2000).  This greatly increases the dynamic range
over which the TF relation can be examined, revealing a number of interesting
points.  a) The ordinary TF relation, plotting
stellar mass in place of luminosity using ${\cal M}_* = \Upsilon_* L$
with $\Upsilon_*^H = 1.0$ and $\Upsilon_*^I = 1.7\;
{\cal M}_{\sun}/L_{\sun}$.  While the usual relation is apparent for massive
galaxies, it breaks down at the low mass end.
b) The gas-only TF relation which follows from
the observed HI (${\cal M}_g = 1.4 {\cal M}_{HI}$), ignoring the stars.
Clearly there is no ``HI TF relation'' for massive galaxies, though there
does seem to be one for low mass objects.
c) The Baryonic TF relation which follows by summing stellar and gas mass:
${\cal M}_d = {\cal M}_* + {\cal M}_g$.  This nicely recovers a
continuous relation over the entire observed range, suggesting that the
disk mass is the fundamental quantity of interest
in the TF relation.  d) The same as (c) but
assuming lower mass-to-light ratios for the stars:
$\Upsilon_*^H = 0.4$ and $\Upsilon_*^I = 0.7$.
This causes a noticeable discontinuity in slope, implying that the
higher mass-to-light ratios adopted in (c) are more appropriate.}
\label{BTF}
\end{figure}

A number of inferences can be drawn from this simple result:
\begin{itemize}
\item Disk mass is the fundamental quantity of interest.
\item Stars got mass!
\item Stars and gas account for nearly all of the disk mass.
\item The product $f_d f_b$ is constant.
\end{itemize}
Items (3) and (4) are just sanity requirements.  If there were another
substantial reservoir of baryons in the disk besides the observed stars
and gas, then there should be some signature of its absence like that
seen in Fig.\ 1(a) \& (b) where an important component has been ignored.
Similarly, the modest scatter in the TF relation only follows if 
$f_d$ is a constant, which only happens naturally if $f_d \approx 1$
(McGaugh et al.\ 2000).  One could consider $f_d \ll 1$ as long as some
mechanism maintained it as a universal constant, or even made it a fine-tuned
function of $V_c$.  Such a situation is highly contrived (van den Bosch
\& Dalcanton 2000).

\section{Stellar Mass-to-Light Ratios}

The baryonic TF relation between disk 
mass and rotation velocity can be expressed as
\begin{equation}
{\cal M}_d = {\cal A} V_c^b.
\end{equation}
A fit to the data (${\cal R} = 0.92$)
in Fig.\ 1(c) has a slope indistinguishable from
$b = 4$ with normalization ${\cal A} \approx 35\,h_{75}^{-2}$.
This line is drawn in each panel of Fig.\ 1.

The greatest source of uncertainty is
the mass-to-light ratios of the stars, which must be assumed.
The stellar mass-to-light ratios assumed in Fig.\ 1(c)
were normalized to the mean of the dynamically determined
$K'$-band maximum disk values for high surface brightness galaxies
(Verheijen 1997) with
colors from stellar population models of de Jong (1996).  These models
also give the same $\Upsilon_*^{K'}$, consistent
with that of the Milky Way (Gerhard, these proceedings) and
the results of Sanders \& Verheijen (1998).

Requiring continuity in the present relation provides an interesting
constraint: stars must have significant mass to avoid the discontinuity
apparent in Fig.\ 1(b) \& (d).  Such mass-to-light ratios are plausible
in terms of both dynamics and stellar populations.  However, this
contradicts the much lower mass-to-light ratios needed for rotation
curve fits with NFW halos and the observed
lack of shift with surface brightness in the TF relation itself.

\section{Conclusions}

The Tully-Fisher relation appears to be a 
manifestation of a more fundamental relation between disk mass and rotation 
velocity (see also de Jong \& Bell, these proceedings).  This relation is
now observed to span over 4 decades in mass,
twice what can be found in most TF studies.  That the TF relation continues
to hold over such a large range, despite the reversal of dominance
of gas at the low masses to stars at the high masses, is a tribute
to its fundamental importance for understanding disk galaxies.  Just why
there should be a relation of the form ${\cal M}_d \propto V_c^4$ remains
open to debate.

\acknowledgments I am grateful to my collaborators, Jim Schombert, Greg
Bothun, and Erwin de Blok for all their many contributions.  I also thank
Ken Freeman, Stephanie Cot\'e, Eric Bell, and Roelof de Jong for stimulating
conversations on this topic.

\end{document}